\documentclass[apj]{emulateapj}
\pdfoutput=1

\lefthead{Giacintucci et al.}
\righthead{A giant radio halo in the massive and merging cluster Abell 1351}

\def\ltapprox{\raise 2pt \hbox {$<$} \kern-1.1em \lower 5pt \hbox {$\approx$}}
\def\ltsim{\raise 2pt \hbox {$<$} \kern-1.1em \lower 4pt \hbox {$\sim$}}
\def\gtsim{\raise 2pt \hbox {$>$} \kern-1.1em \lower 4pt \hbox {$\sim$}}

\def\ltsim{\raise 2pt \hbox {$<$} \kern-0.8em \lower 4pt \hbox {$\sim$}}
\def\gtsim{\raise 2pt \hbox {$>$} \kern-0.8em \lower 4pt \hbox {$\sim$}}

\def\eg{{\it e.g.,~}}

\shorttitle{A giant radio halo in Abell 1351}
\shortauthors{Giacintucci et al.}

\begin{document}

\title{A giant radio halo in the massive and merging cluster Abell 1351}

\author{S. Giacintucci\altaffilmark{1,2} 
T. Venturi\altaffilmark{2},
R. Cassano\altaffilmark{2,3},
D. Dallacasa\altaffilmark{2,3},
G. Brunetti\altaffilmark{2}}

\altaffiltext{1}{Harvard-Smithsonian Center for Astrophysics, 60 Garden
Street, Cambridge, MA 02138, USA}
\altaffiltext{2}{INAF/IRA, via Gobetti 101,
I--40129 Bologna, Italy}
\altaffiltext{3}{Dip. Astronomia, University of Bologna, via Ranzani 1,
I--40127 Bologna, Italy}


\begin{abstract}
We report on the detection of diffuse radio emission in the X-ray 
luminous and massive galaxy cluster A\,1351 (z=0.322) using archival Very Large Array data at 1.4 GHz. Given its central location, morphology, and Mpc-scale extent, we classify the diffuse source as a giant radio halo. X-ray and weak lensing studies show A\,1351 to be a system undergoing a major merger. The halo is associated with the most massive substructure. The presence of this source is explained assuming that merger-driven turbulence may re-accelerate high-energy particles in the intracluster medium and generate diffuse radio emission on the cluster scale. The position of A\,1351 in the logP$_{\rm 1.4~GHz}$ -- logL$_{\rm X}$ plane is consistent with that of all other radio-halo clusters known to date, supporting a causal connection between the unrelaxed dynamical state of massive ($>10^{15}$ $M_{\odot}$) 
clusters and the presence of giant radio halos. 
\end{abstract}



\section{Introduction}

Beyond the galaxies and the hot gas, clusters of galaxies host non-thermal 
components in the form of $\mu$G magnetic fields and relativistic particles 
mixed with the thermal intracluster medium (ICM), which prove themselves through steep-spectrum diffuse radio emission in the form of radio halos and relics (see Ferrari et al. (2008) for a recent review). These powerful sources (radio power P$_{\rm 1.4 \, GHz} \sim$ 10$^{24}$--10$^{25}$ W Hz$^{-1}$) are extended on the Mpc-scale and are characterized by very low surface brightness ($\mu$Jy arcsec$^{-2}$ at 1.4 GHz). The origin of these sources has long been considered a puzzle, due to their rare occurrence and to the unclear origin of the radiating relativistic particles (\eg Brunetti (2008) and Cassano (2009) for recent reviews).

Radio halos are always found in unrelaxed clusters (Buote 2001; Venturi
et al. 2008) showing significant substructure in the X-ray images, as well 
as complex gas temperature distribution (\eg Govoni et al. 2004; Giacintucci et al. 2005, 2009), which are clear signatures of very recent or ongoing merger events. This suggests a link between the gravitational process of cluster formation and the origin of radio halos. The presence of a correlation between the radio power of halos and the X-ray luminosity, mass, and temperature of the host cluster further supports a tight connection between the thermal and non-thermal components in the ICM (Liang et al. 2000; Govoni et al. 2001; Cassano et al. 2006, hereinafter C06). 
Recent work based on the Giant Metrewave Radio Telescope (GMRT) {\it radio 
halo survey} (Venturi et al. 2007, 2008; hereinafter V07, V08), carried out 
with the GMRT at 610 MHz, firmly established that (1) radio 
halos are not ubiquitous in clusters, with only $\sim30\%$ of the selected 
X-ray luminous clusters hosting a radio halo (V08; Cassano et al. 2008, 
hereinafter C08), and (2) the distribution of radio halos in the 
P$_{\rm 1.4 \, GHz}$ -- cluster X-ray luminosity (L$_{\rm X}$) plane is 
{\it bimodal}: clusters with and without radio halos are clearly separated, 
with radio halos always found in dynamically disturbed systems (Brunetti et al. 
2007; V08). 

A model that explains the origin of radio halos is provided by the so-called 
{\it turbulent re-acceleration} scenario (Brunetti et al. 2001; Petrosian 2001), where it is assumed that magnetohydrodynamic (MHD) turbulence, injected in the ICM during cluster--cluster mergers, has an important role in the re-acceleration of relativistic electrons. According to this scenario, radio halos are expected to be relatively {\it rare} 
and {\it transient} phenomena (with a lifetime $\leq 1$ Gyr) generated in dynamically active clusters. The fraction of galaxy clusters with radio halos is expected to increase with cluster mass, since mergers in massive clusters are expected to generate enough turbulence to accelerate relativistic electrons at energies of several GeV (Cassano \& Brunetti 2005).

Based on the above arguments, and motivated by the weak lensing analysis
of the massive and unrelaxed cluster A\,1351 (Holhjem et al. 2009, hereinafter H09, and references therein), not known to host a radio halo, we searched the Very Large Array (VLA) data archive to check for possible hints of diffuse radio emission on the cluster scale.

In this Letter, we report on the detection of a giant radio halo at the center of A\,1351 using archival VLA data at 1.4 GHz. The radio observations and data reduction are described in Section \ref{sec:obs}, 
the radio images and their analysis are reported in Section \ref{sec:results}; 
results are presented and discussed in Section \ref{sec:disc}; summary and 
conclusions are given in Section \ref{sec:summ}.

We adopt the $\Lambda$CDM cosmology with H$_0$=70 km s$^{-1}$ Mpc$^{-1}$, 
$\Omega_m=0.3$ and $\Omega_{\Lambda}=0.7$. At the redshift of 
A\,1351 (z=0.322) this leads to a linear scale of 
$1^{\prime \prime}=4.68$ kpc. 

\section{A\,1351 and the radio data}\label{sec:obs}

A\,1351 is a rich galaxy cluster at redshift z=0.322 with an
X-ray luminosity of L$_{\rm X\,[0.1-2.4]\, keV}= 8.4\times10^{44}\,
\mathrm{h_{70}}^{-2}$ erg s$^{-1}$ (Allen et al. 2003) and a total mass 
of $M_{200}\simeq1.17 \pm 0.31 \times 10^{15} h_{70}^{-1} M_{\odot}$ (H09). 
X--ray images show significant substructure, 
suggesting an ongoing major merger (Allen et al. 2003). The cluster has a high 
optical velocity dispersion ($\sigma_{\rm v} >$1000 km s$^{-1}$; H09 and references 
therein), as commonly found in merging systems. The merging scenario 
is supported by the weak lensing analysis presented in H09: two 
significant mass concentrations are detected in the weak lensing reconstruction 
of the cluster mass density, coincident with the galaxy and X--ray gas distribution. 
The two mass substructures are located at a 
projected distance of $\sim$ 3$^{\prime}$ ($\sim 840$ kpc) along the 
northeast/southwest axis, suggesting that this is the direction of the merger.
\\
\\
A\,1351 was observed with the VLA at 1.4 GHz in the A-, D- and 
C-array configurations in 1994 April, 1995 March and 2000 April,
respectively. The details on these observations are summarized in 
Table \ref{tab:obs}, where the columns provide the following 
information: VLA project, frequency, total bandwidth, array configuration,
total time on source, observing date, half-power bandwidth (HPBW), and
rms level (1$\sigma$) in the full resolution image from each individual 
data set.


\begin{table*}[t]
\caption[]{Details of the VLA archival observations}
\begin{center}
\begin{tabular}{cccccccc}
\hline\noalign{\smallskip}
VLA     & Frequency & $\Delta \nu$ & Config. & Obs. Time & Date & Beam, P.A.  &   rms      \\  
Project &   (GHz)   &      (MHz)   &         &  (minutes)    &       & ($^{\prime \prime} 
\times^{\prime \prime}$, $^{\circ}$) & ($\mu$Jy b$^{-1}$) \\
\noalign{\smallskip}
\hline\noalign{\smallskip}
AB699   &    1.4    &   25 &  A  & 30    & 1994 Apr & 1.5 $\times$ 1.1, $-$5 &   65 \\ 
AO149   &    1.4    &   25 &  C  & 120   & 2000 Apr & 15.9 $\times$ 13.3, 28 &   50 \\
AM469   &    1.4    &   25 &  D  & 60    & 1995 Mar & 57.9 $\times$ 43.0, 50 &   200 \\
\hline{\smallskip}
\end{tabular}
\end{center}
\label{tab:obs}
\end{table*}


The data were calibrated and reduced using the Astronomical Image Processing 
System (AIPS), following the standard procedure. Each data set was 
self-calibrated in phase to correct for residual phase variations.  
Both uniform and natural weighting were used to produce the final images
for each array configuration. The achieved sensitivity (1$\sigma$) ranges 
from 50 to 200 $\mu$Jy b$^{-1}$ (see Table \ref{tab:obs}), the spread being 
mostly due to the different observing time. The final self-calibrated data sets 
were then combined together in a single data set providing a UV range of 
0.2-175 k$\lambda$. The shortest baselines ensure the proper imaging of extended 
radio emission up to $\sim$7$^\prime$.5 in angular size, corresponding to 
$\sim$ 2.1 Mpc at the redshift of A\,1351. The high angular resolution 
provided by the longest baselines allows us to identify and separate
possible discrete radio sources projected onto the diffuse emission.
With the final combined array we made a set of images, with resolutions 
ranging from $6^{\prime\prime}.7 \times 5^{\prime\prime}$.5 (full resolution, 
uniform weighting and 1$\sigma=30$ $\mu$Jy b$^{-1}$) to $20^{\prime\prime}.2 
\times 17^{\prime\prime}.8$ (natural weighting, taper, and 1$\sigma=60$ $\mu$Jy 
b$^{-1}$). The average residual amplitude errors in the data are on
the order of $\ltsim$ 5\%.

\begin{figure*}[t]
\epsscale{1.15}
\plotone{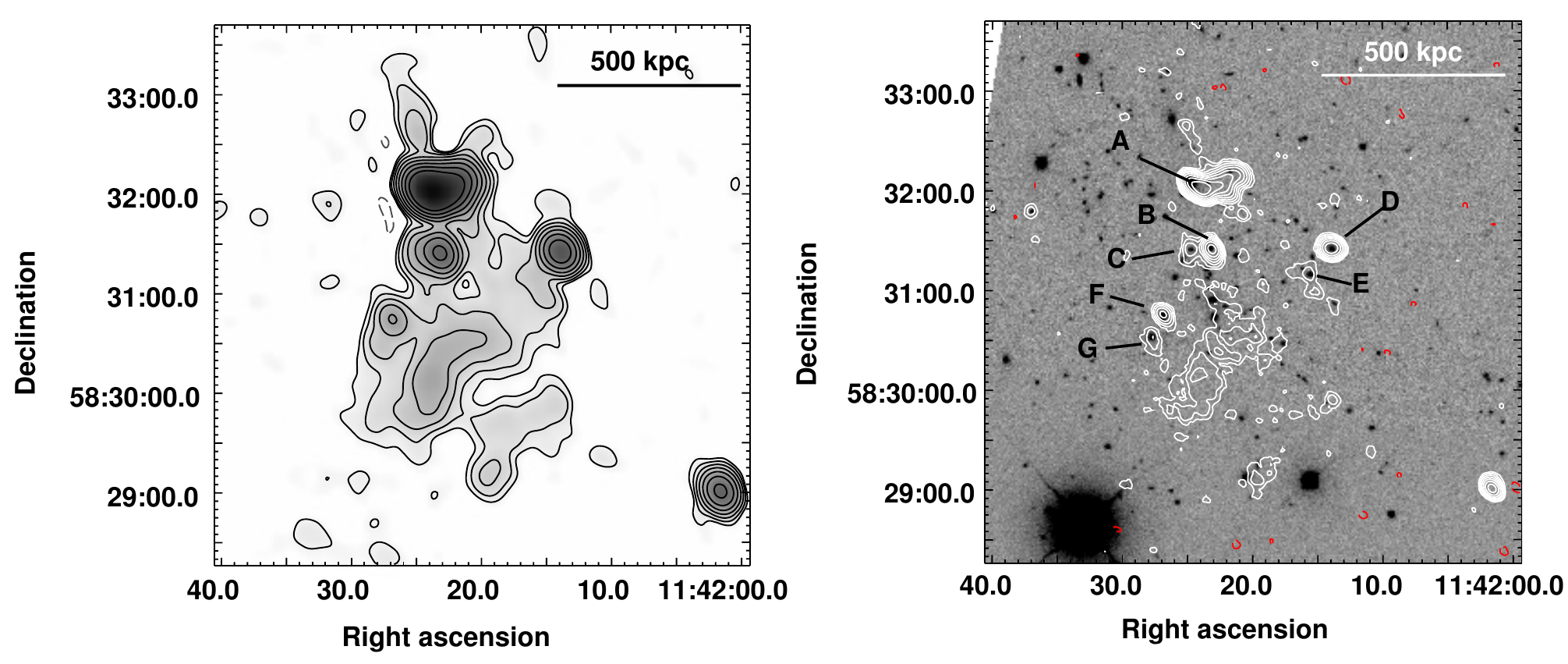}
\caption[]{{\it Left:} VLA 1.4 GHz image (contours and grayscale) obtained
from the combination of the C-- and D--array data sets. The HPBW is 
$16^{\prime\prime}.2 \times 13^{\prime\prime}.6$, P.A. 28$^{\circ}$.
Contours are spaced by a factor 2 from $\pm$3$\sigma$=0.15 mJy b$^{-1}$. 
{\it Right:}  VLA 1.4 GHz contours from the combined (A+C+D)--array data set, 
superposed to the red optical image from SDSS. The HPBW is $6^{\prime\prime}.7 
\times 5^{\prime\prime}.5$, P.A. 40$^{\circ}$. Contours are spaced by a factor 
2 from $\pm$3$\sigma$=0.09 mJy b$^{-1}$ (negative contours are shown in red). 
Individual radio galaxies are labeled from A to G.} 
\label{fig:radio_opt}
\end{figure*}

\section{The radio images}\label{sec:results}

The left panel of Figure \ref{fig:radio_opt} shows the 1.4 GHz image of the
radio emission from the cluster, at the resolution of $16^{\prime\prime}.2 
\times 13^{\prime\prime}.6$, obtained from the combination of the C- and
D-array data sets. Diffuse emission is clearly present beyond the individual 
sources.
The right panel of Figure 1 shows a $6^{\prime\prime}.7 
\times 5^{\prime\prime}.5$ image obtained with the combined arrays, overlaid 
on the Sloan Digital Sky Survey (SDSS) red optical image, to highlight the 
optically identified radio sources (labeled from A to G) detected in the 
cluster region. Source A is extended 
and is associated with SDSS J114224.77+583205.3, the 
brightest cluster galaxy (BCG). The diffuse emission has no obvious optical 
counterpart, and permeates the cluster region south of the BCG.

\begin{figure}
\epsscale{1.1}
\plotone{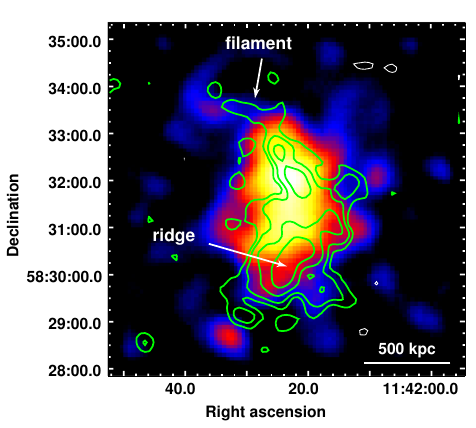}
\caption[]{VLA 1.4 GHz contours of the extended emission in A\,1351, after 
subtraction of the individual sources (A to G in Figure \ref{fig:radio_opt};
left panel) from the combined (A+C+D)--array data set. The image has been 
obtained using the UV range 0.2--16 k$\lambda$. The HPBW is 
$20^{\prime\prime}.2 \times 17^{\prime\prime}.8$, P.A. 28$^{\circ}$. Contours 
are spaced by a factor 2, starting from $\pm$3$\sigma$=0.18 mJy b$^{-1}$ 
(negative contours are shown in white). The color image is the smoothed 
{\it ROSAT} HRI image of the cluster.}
\label{fig:halo}
\end{figure}

Figure \ref{fig:halo} presents a low-resolution image of the radio emission
at the cluster center, overlaid on the smoothed X-ray image obtained from 
archival {\it ROSAT} High Resolution Imager (HRI) data. The radio image was 
produced after subtraction of the discrete radio galaxies  (A to G 
in Figure \ref{fig:radio_opt}), and using the UV range 0.2--16 k$\lambda$. 
The image confirms the presence of a diffuse radio source extending on the 
cluster scale, whose elongation and overall size are in good agreement with 
the brightest region detected in the X-ray. The radio morphology is 
irregular, with a peak of emission coincident with the X-ray peak. 
A southern feature, which we label {\it ridge}, is spatially coincident 
with an edge in the X-ray image. In the northern region a faint filament 
of radio emission extends toward the northeast, similar to the X-ray emission. 
Given its location, morphology, and largest linear size of $\sim 1.1$ Mpc, we 
classify the diffuse source in A\,1351 as a {\it giant radio halo}. The 
{\it ridge} is the brightest and most peripheral region of the halo, and 
its location with respect to the X-ray brightness distribution raises the 
possibility that it might be a radio relic projected onto the radio halo 
emission.
\\
Comparison between the morphology of the radio halo (Figure \ref{fig:halo}) and 
the mass reconstruction in H09 (their Figure 5) shows that the halo is associated 
with the most massive cluster substructure, and the northern radio filament 
(Figure \ref{fig:halo}) 
points toward a secondary peak both in the mass reconstruction and number 
density of galaxies.
\\
The total flux density of the radio halo in Figure \ref{fig:halo} is 
S$_{\rm 1.4 \, GHz}$= 32.4 mJy, corresponding to a radio power P$_{\rm 1.4 
\, GHz}$= 1.17$\times$10$^{25}$ W Hz$^{-1}$. The {\it ridge} accounts for approximately 
40\% of the total flux density (P$_{\rm 1.4 \, GHz}$= 4.67$\times$10$^{24}$ W Hz$^{-1}$).
We note that the contribution of residual emission from the head--tail 
(source A in Figure \ref{fig:radio_opt}) is $\ltsim$ 4 mJy, as estimated from 
the comparison of the flux measurements on a set of different resolution 
images of the whole radio emission from the cluster (including the diffuse 
emission and the contribution of the individual radio galaxies). 

\section{Discussion}
\label{sec:disc}

The current picture of radio halo formation in galaxy clusters 
is very complex and tightly connected to the process of formation 
and evolution of the host cluster (\eg Brunetti (2008) for a recent
review). Monte Carlo simulations, carried out under the hypothesis that 
MHD turbulence in cluster mergers may re-accelerate high-energy 
particles, show that about 30\% of X-ray luminous (L$_{\rm X} > 
8 \times 10^{44}$ erg s$^{-1}$) galaxy clusters at z$\sim$ 0.2-0.3 
are expected to host a giant radio halo (C06; C08). In this context, 
the presence of diffuse cluster-scale radio emission in 
A\,1351 is not surprising since the cluster is a massive and X-ray 
luminous system experiencing a major merger event (Allen et al. 2003; 
H09). In particular, Cassano \& Brunetti (2005) derived 
an analytical formula for the turbulent-acceleration rate based on a 
single-merger scenario. Assuming a redshift z=0.322, we find that 
steep-spectrum Mpc-scale radio emission at 1.4 GHz (that requires 
acceleration timescales $\leq 0.3$ Gyr) is expected to be generated in 
connection with a merger with mass ratio $\leq 2.8$, for a total virial
mass of the system $M_v \simeq 1.2 \times 10^{15} M_{\odot}$ 
(consistent with the weak lensing analysis in H09), or $\leq 4$, 
for $M_v \simeq 2 \times 10^{15} M_{\odot}$ (derived from the 
$L_x$--$M_v$ correlation; \eg C06).
Furthermore, based on the GMRT observations in V07 and V08 and 
literature data, C08 showed that the fraction of 
clusters with radio halo in the redshift range $0.20-0.32$ 
and with $L_X>8 \times 10^{44}$ erg s$^{-1}$ is on the order 
of $40\%$, in line with the expectations of the re-acceleration scenario.
Remarkably, the X-ray luminosity ($L_X \simeq 8.4 \times 10^{44}$ erg 
s$^{-1}$) and redshift (z$=0.32$) of A\,1351 are within these 
same ranges.

\begin{figure}[h!]
\plotone{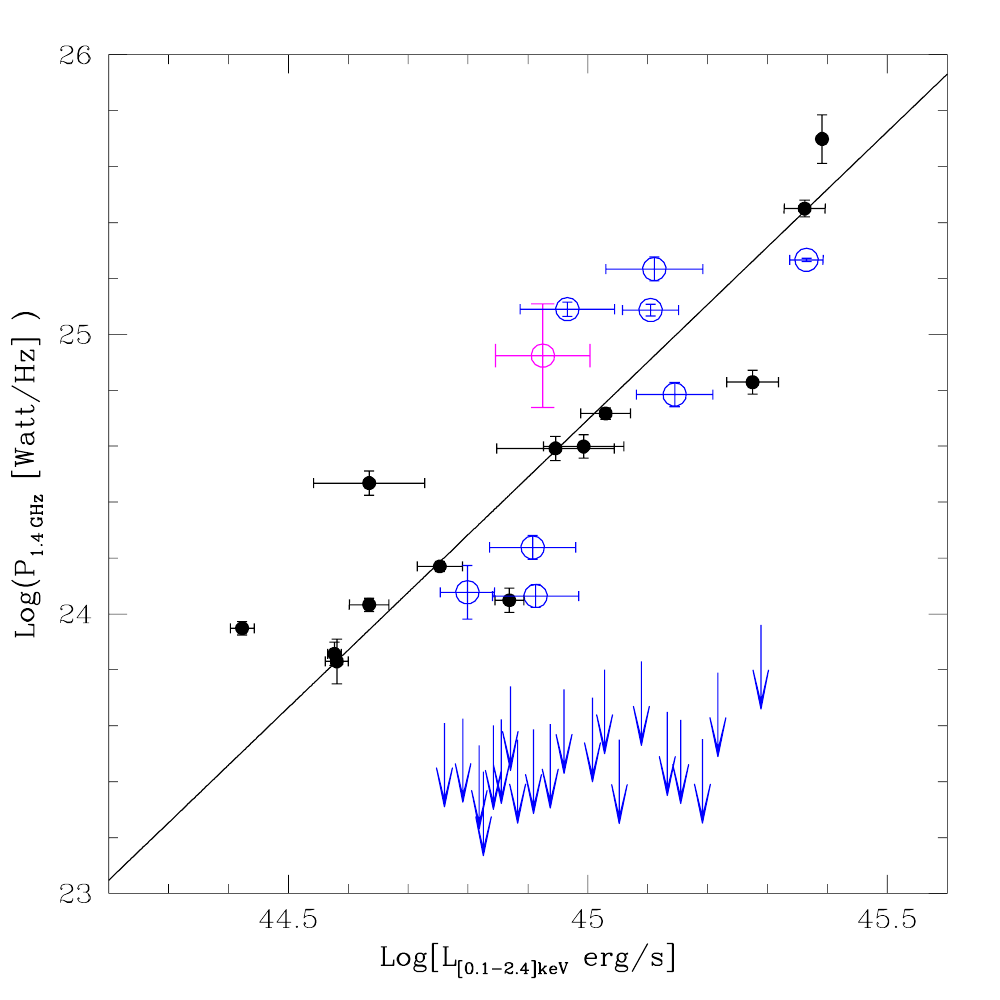}
\caption[]{Distribution of the GMRT radio-halo clusters (open blue symbols) and 
other radio-halo clusters from the literature (filled black symbols) in the 
$P_{1.4 \, GHz}-L_X$ plane (adapted from Brunetti et al. 2009). The magenta point marks 
the position of A\,1351.}
\label{fig:corr}
\end{figure}

In Figure \ref{fig:corr} we show the distribution of the clusters of the 
{\it GMRT radio halo survey} and of other radio-halo clusters from the literature 
in the $P_{1.4\, GHz}-L_X$ plane (adapted from Brunetti et al. 2009).
The clusters hosting a radio halo are clearly separated from those
without. The analysis of the dynamical properties (where available) 
suggests that such separation is most likely due to the different dynamical 
state of {\it radio-loud} and {\it radio-quiet} clusters: radio halos are 
always found in dynamically disturbed clusters, while {\it radio-quiet} 
clusters tend to be relaxed systems (V08). This can be interpreted 
in the framework of the re-acceleration scenario since clusters are expected 
to move with time from the {\it radio-quiet} region (upper limits in 
Figure \ref{fig:corr}) to the P$_{1.4 \, GHz}-L_X$ correlation in a relatively 
short timescale (on the order of $\approx 10^8$ years, i.e., 
the re-acceleration timescale of the emitting electrons), leaving the 
region between radio-halo and {\it radio-quiet} clusters poorly populated
(Brunetti et al. 2007, 2009). 

In Figure \ref{fig:corr} we report the position of A\,1351 as a magenta symbol. 
The diffuse radio emission in A\,1351 has a complex morphology, and 
the data currently available do not allow us to establish if the 
{\it ridge} (Figure \ref{fig:halo}) is part of the radio halo emission or  
a radio relic located in the southern region of the cluster. 
For this reason, the logP$_{\rm 1.4 \,GHz}$ value adopted in Figure \ref{fig:corr}
(logP$_{\rm 1.4 \, GHz}$ [W Hz$^{-1}$]$=24.97\pm0.19$) is an intermediate value, 
whose associated large errors account for the possibility that the {\it ridge} is 
either part of the radio halo or unrelated to it. The location of A\,1351 is consistent 
with the distribution of all radio halos known so far. We underline that the plot 
in Figure \ref{fig:corr} includes all the information available in the literature 
so far: only $\sim$ 20 clusters are known to host a giant radio halo, therefore
the detection of a new one in a well-constrained merging system contributes to 
settle the radio-halo cluster merger connection.

\section{Conclusions}\label{sec:summ}

Using archival VLA data at 1.4 GHz we found a giant radio halo
in the massive galaxy cluster A\,1351, with a total extent of $\sim$ 1.1 
Mpc and a radio power P$_{\rm 1.4 \, GHz}$= 1.2$\times$10$^{25}$ W Hz$^{-1}$.
The host cluster is experiencing a major merger process which is responsible 
for the complex distribution of the X-ray emitting gas and total mass of the 
system, as revealed by the X-ray and weak lensing studies. 
The ongoing merger may generate turbulence in the cluster volume, which in 
turn may re-accelerate relativistic electrons in the ICM and generate Mpc-scale
radio emission. 
Monte Carlo simulations, carried out under this hypothesis, show that about 
30$\%$ of X-ray luminous (L$_{\rm X} > 8\times 10^{44}$ erg s$^{-1}$) 
galaxy clusters at z$\sim$ 0.2-0.3 are expected to host a giant radio halo 
(C06, C08).
Remarkably, the position of A\,1351 in the 
plane P$_{\rm 1.4 \, GHz}$--$L_{\rm X}$ is consistent with that of all the 
other radio-halo clusters known to date, suggesting a causal link between the 
dynamical state of massive ($>10^{15}$ $M_{\odot}$) clusters and the presence 
of giant radio halos. 

Follow-up observations at lower (and multi) frequency will be important 
to derive the total radio spectrum of the diffuse emission, which is an
essential information to discriminate between the theoretical models for 
the formation of radio halos (e.g., Brunetti et al. 2008, Dallacasa et al.
2009). A spatially resolved 
image of the spectral index distribution in the halo will also be a
useful tool to shed light on the origin of the diffuse radio component
in A\,1351.

\section{Acknowledgements}
This work is partially supported by INAF under grants PRIN-INAF2007 and 
PRIN-INAF2008 and by ASI-INAF under grant I/088/06/0.

\end{document}